\documentstyle[sprocl]{article}

\def\Journal#1#2#3#4{{#1} {\bf #2}, #3 (#4)}

\def\NPB{{\em Nucl. Phys.} B}
\def\PLB{{\em Phys. Lett.}  B}

\def\PRD{{\em Phys. Rev.} D}

\def\be{\begin{equation}}
\def\ee{\end{equation}}
\def\bea{\begin{eqnarray}}
\def\eea{\end{eqnarray}}

\begin{document}

 	\begin{flushright}
  		 CTP-TAMU-48/98\\
                 SINP-TNP/98-30\\
                 hep-th/9812011\\
     \end{flushright}
     
\vskip .5cm

\title{ ON THE CONSTRUCTION OF SL(2,Z)  TYPE IIB 5-BRANES}

\author{J. X. LU}

\address{Center for Theoretical Physics\\
Department of Physics, Texas A \& M University\\
College Station,  TX 77843, USA\\E-mail: jxlu@rainbow.physics.tamu.edu}

\author{Shibaji Roy}

\address{Saha Institute of Nuclear Physics\\
1/AF Bidhannagar, Calcutta 700 064, India\\E-mail: roy@tnp.saha.ernet.in}


\maketitle\abstracts{This talk reviews our recent work on the construction of 
SL(2,Z) multiplets of type IIB superfivebranes. We here pay 
particular attention 
to the methods employed and some salient features of the solutions.}

\section{Introduction}
The study of perturbative superstring spectra indicates that there must exist
a so-called Type IIB supergravity as the low energy limit of type IIB
superstring in addition to the well-known type IIA supergravity. This theory has
two spacetime supersymmetries with the same handedness and is therefore chiral.
The field
content for this theory (in the real basis) consists of the metric 
$g_{\mu\nu}$, a 2-from gauge
potential $B^{(1)}_{\mu\nu}$, and the dilaton $\phi$ as its bosonic fields in 
the
so-called NSNS-sector. It also contains another scalar $\chi$, another 
2-form gauge potential 
$B^{(2)}_{\mu\nu}$ and a 4-form gauge potential $A_{\mu\nu\rho\sigma}$ 
as the
bosonic fields in the so-called RR-sector. In the fermionic 
sector the spectrum
consists of two left-handed Majorana-Weyl
gravitinos $\psi^{(i)}_\mu$ and two right-handed Majorana-Weyl spinors 
$\lambda^{i}$ with ($i =1, 2$).  The field strength for the 4-form gauge
potential $A_{\mu\nu\rho\sigma}$ is self-dual. For this reason, the
construction of a manifestly
covariant action is not possible for this theory.

This fact led Schwarz \cite{sch} to write down the covariant equations
 of
 motion instead
 for this theory. The key for doing this is the supersymmetry and the
 global SL(2,R) symmetry. 
The two supersymmetries in this theory are known from the 
 outset. However,
 the global SL(2,R) symmetry is not so obvious. But its existence can be 
inferred 
 from the following.
 We have two supersymmetries here which 
 can be related to each other by a SO(2) rotation. In other words,  
the superalgebra 
 has SO(2) as its automorphism
 group. The past experience on Cremmer-Julia hidden symmetries tells us 
that type IIB 
 supergravity might have 
 some global non-compact group as its symmetry group whose maximum 
compact subgroup is $SO(2)$. 
 That there are two scalars in this
 theory tells us that the possible global symmetry is SL(2,R). 
The scalars should parametrize
 the coset SL(2,R)/SO(2). The two 2-form potentials transform under 
the SL(2,R) as a doublet
 while the remaining fields are inert under the SL(2,R) when 
the metric is written in Einstein 
 frame. So the corresponding charge doublet whether electric-like 
(strings) or magnetic-like (fivebranes)
 will also transform as a doublet accordingly.
  
  The existence of this global SL(2,R) symmetry in type IIB supergravity is 
  not consistent with the perturbative type IIB
  superstring. For example, its maximum compact subgroup SO(2) is a 
symmetry of the 
  corresponding superalgebra but not
  a symmetry of type IIB superstring worldsheet action \cite{tow}. Thanks 
to the recent
  understanding of non-perturbative string theory, 
we now know that only the discrete
  subgroup SL(2,Z) will survive quantum mechanically as a symmetry of the
  non-perturbative type IIB string theory. In order for this to be true,
  non-perturbative extended objects like D-string and 
D5-brane both carrying RR charges
  and NSNS 5-brane carrying NSNS charge must be included in the 
non-perturbative 
  spectrum. As we will see, precisely the requirement of charge quantization
  break the continuous or classical SL(2, R) symmetry to the discrete 
SL(2,Z).  Except
  for the cases involving dyonic objects, the same applies to 
various U-duality symmetries 
  of M-theory from the Cremmer-Julia continuous correspondences.

  Unlike all the other string dualities in $D = 10$, 
the type IIB SL(2,Z) is indeed 
  a symmetry of type IIB superstring theory. In particular, 
this SL(2,Z) contains a
  transformation which maps the dilaton $\phi \rightarrow - \phi$, 
therefore is a strong-weak
  duality symmetry. In this respect, it is like the 
S-duality SL(2,Z) of $N = 4,  D = 4$
  heterotic string \cite{rey,fon,sen}. In other words, type IIB string 
theory is self-dual. 
  However, unlike the heterotic SL(2,Z) which rotates 
electrically and magnetically charged
  states of the same gauge field, the present SL(2,Z) 
relates either electrically or
  magnetically charged states of {\it two} different gauge fields. 
This makes it more like a
  T-duality \cite{schone}. 
  
  What can we make use of the quantum SL(2,Z) symmetry of 
the non-perturbative type IIB
  string theory? To our knowledge, new information can be 
learned based on this quantum SL(2,Z)
  symmetry in the following three cases: 1) The most general 
stable BPS electrically charged 
  superstring states \cite{schone} or magnetically charged 
superfivebrane states \cite{lur}
  can be constructed from the corresponding known 
single-charge solution based on this symmetry.
  2) Certain SL(2,Z) invariant terms in the effective action of 
type IIB string theory can be 
  deduced from the symmetry requirement and the explicitly known  
perturbative terms 
  (with possible application of some known perturbative 
  non-renormalization theorem) as reviewed recently by 
Sen \cite{senone} and references therein.
  3)This symmetry is not only the basis for the 
AdS/CFT correspondence of Type IIB string on
  $AdS_5 \times S^5$ and $N = 4, D =4$ super Yang-Mills on the 
boundary for both $g_s > 1$ and $g_s < 1$, 
  with $g_s$ the type IIB string coupling, but also is an 
important ingredient for this correspondence to be 
  true.

  In this brief review, we only focus on how to construct, 
in particular, the most general stable BPS
  fivebranes carrying magnetic charges from a given simple 
fivebrane solution in type IIB string theory
   in case 1) above. We also discuss various salient features 
in this construction.

\section{General Stable BPS Fivebranes}
To construct SL(2,Z) type IIB fivebranes carrying magnetic charges, we as usual start with
a known pure NSNS superfivebrane \cite{dufl} with zero asymptotic values 
of the scalars 
in type IIB theory.  Depending on the charge carried by the NSNS fivebrane to 
be a quantized unit one or just an
arbitrary classical one, there exist basically two methods which can be used to construct the SL(2,Z)
fivebranes. In the former case, a compensating factor needs to be introduced to the initial unit charge by hand
such that the transformed charge doublet, obtained after the classical 
SL(2,R) transformation  
can remain to be quantized (this is the method used by Schwarz
 \cite{schone} to construct
the SL(2,Z) strings). By this, both the compensating factor and the SL(2,R)
matrix are completely determined in terms of  the vacuum moduli and the quantized charges. In the latter case,
an initial charge doublet with the arbitrary classical NSNS charge as its only non-vanishing componenet is
transformed by a SL(2,R) transformation to a general but given classical 
charge doublet. By this, both the initial NSNS charge and the elements of
SL(2,R) matrix are completely determined in terms of the asymptotic values 
of scalars and the given classical charges. Then we impose the charge
quantization on each component of the given charge doublet due to the 
existence of F-string and D-string, 
the magnetic duals of
NSNS fivebrane and RR fivebrane, respectively. The two methods produce 
the same general SL(2,Z) fivebrane solution 
but they have different
implications. For the former method, we sandwich a classical SL(2,R) 
transformation between quantum
mechanically allowable initial and final fivebrane configurations. As a 
consequence, the mass of the final 
configuration is different from that of the initial configuration 
by the compensating factor introduced by
hand while a SL(2,R) transformation preserves the mass. We do not have 
such a problem in the second method. We
therefore employ it here to construct the general SL(2,Z) 
superfivebranes. But before doing so, let us review briefly
the relevant part of type IIB supergravity and a pure NSNS BPS fivebrane 
configuration.

Since we are interested in vacuum-like BPS solutions, the fermionic fields 
are set to zero from the outset. 
Further, dropping the irrelevant 4-form potential, we can write the 
corresponding action of type IIB supergravity
in a SL(2,R) invariant form \cite{schone,hul} (in Einstein frame) as 
\begin{equation}
S = {1\over 2 \kappa^2} \int d^{10}x  \sqrt{- g}   \bigg[R  + \frac{1}{4} tr \nabla_\mu {\cal M} \nabla^\mu 
{\cal M}^{-1} - \frac{1}{12} {\cal H}_{\mu\nu\rho}^T 
{\cal M} H^{\mu\nu\rho} \bigg].
\label{eq:ia}
\end{equation}
In the above, the two scalars $\chi$ and $\phi$ parametrize the coset 
SL(2,R)/SO(2) as
\begin{equation}
{\cal M}\,\, = \,\, \left(\begin{array}{cc}
\chi^2 + 
e^{- 2{\phi}}
&  \chi \\
 \chi  &  1\end{array}\right)\,\,e^{{\phi}},
\label{eq: matrix}
\end{equation}
and the NSNS 3-form field strength $H^{(1)} = d B^{(1)}$ is combined 
with the RR 3-form field strength 
$H^{(2)} = d B^{(2)}$ to form a doublet as
\begin{equation}
{\cal H}_{\mu\nu\lambda} = \left(\begin{array}{c}
H_{\mu\nu\lambda}^{(1)} \\ H_{\mu\nu\lambda}^{(2)}\end{array}\right).
\label{eq:hd}
\end{equation}

The action~(\ref{eq:ia}) can be easily seen to be invariant under the following global SL(2,R) transformation:
\begin{equation}
 {\cal M} \rightarrow \Lambda {\cal M} \Lambda^T, \qquad 
{\cal H}_{\mu\nu\lambda} \rightarrow \left(\Lambda^{-1}\right)^T {\cal H}_{
\mu\nu\lambda}, \qquad g_{\mu\nu} \rightarrow g_{\mu\nu},
\label{eq:transf}
\end{equation}
where $\Lambda = \left(\begin{array}{cc} a & b\\ c & d\end{array}\right)$, with
$ad - bc = 1$, represents a global SL(2, R) transformation matrix. 
It is
easy to check that under the transformation~(\ref{eq:transf}), the complex scalar
$\lambda = \chi + i e^{-\phi}$ transforms as,
\begin{equation}
\lambda \rightarrow \frac{a\lambda + b}{c\lambda + d}.
\label{eq:lamb}
\end{equation}

We would like to point out here that unlike the electrically charged
string solution, the magnetic charges associated with 
$H_{\mu\nu\lambda}^{(1)}$ 
and $H_{\mu\nu\lambda}^{(2)}$ of the five-brane should transform in the
same way as the field strengths themselves. This follows from the fact
that Noether charge (or the electrical charge) of the string solution
is conserved due to the equation of motion following from~(\ref{eq:ia})
whereas the topological charge (or the magnetic charge) of the five-brane
is conserved due to Bianchi identity. Therefore the magnetic charges
of the five-branes transform as ${\cal P} \rightarrow 
(\Lambda^{-1})^T {\cal P}$        
or
in components,
\begin{equation}
\begin{array}{rcl}
P^{(1)} &\rightarrow& d P^{(1)} - c P^{(2)}\\
P^{(2)} &\rightarrow& -b P^{(1)} + a P^{(2)}
\end{array}
\label{eq:charget}
\end{equation}

Now in order to construct the general BPS superfivebrane configuration 
all we need is a known
particular superfivebrane solution and the condition that both  
$P^{(1)}$ and $P^{(2)}$ should
be quantized separately. The pure NSNS superfivebrane configuration has been 
found some time ago~\cite{dufl}
from an action involving pure NSNS fields which can be obtained from the action~(\ref{eq:ia}) when the RR fields 
$\chi$ and $H^{(2)}_{\mu\nu\rho}$ are consistently set to zero. 
This configuration preserves one half of the
spacetime supersymmetries. With zero asymptotic value of the dilaton 
and insisting that the metric approach Minkowski spacetime asymptotically, 
we have the following 
superfivebrane configuration~\cite{dufl}
\begin{equation}
\begin{array}{rcl}
ds^2 &=& \left(1 + \frac{P}{2 \rho^2}\right)^{-1/4}\left[-(dt)^2 +
\delta_{ij} dx^i dx^j\right]
+\left(1 + \frac{P}{2\rho^2}\right)^{3/4}\left(d\rho^2 + \rho^2 
d\Omega^2_3\right)\\ 
 e^{2\phi}\,\,\,& = &\,\,\, \left(1 + 
\frac{P}{2 \rho^2}
\right),\qquad H^{(1)}\,\,\,=\,\,\,P \epsilon_3.
\end{array}
\label{eq:nsfiveb}
\end{equation}
Here $i, j = 1, 2, 3, 4, 5$ and $d\Omega^2_3$ is the metric on the unit 
3-sphere. $\epsilon_3$ is the
corresponding volume form. At this point, the charge $P$ is an 
arbitrary classical one.

We have three steps to  construct a general stable SL(2,Z) superfivebrane configuration using the second method described
above. The first step consists of the construction of a general {\it classical}
superfivebrane  with non-vanishing but fixed 
asymptotic value of the complex scalar $\lambda_0 = \chi_0 + i e^{-\phi_0}$ in the sense that the charges 
in the charge doublet are classical. We first seek a most general SL(2,R) transformation $\Lambda$ such that it maps the 
initial asymptotic value of the complex scalar $\lambda_0 = i$ to the given $\lambda_0 = \chi_0 + i e^{-\phi_0}$.
Here we use the subscript ``0" to denote the asymptotic values of 
the scalars. For example, 
$\phi_0$ denotes the asymptotic value of the dilaton. By this, 
the SL(2,R) matrix $\Lambda$ can be partially determined as
\begin{equation}
\Lambda =
\left(\begin{array}{cc} 
e^{-\phi_0} \cos \alpha + \chi_0 \sin \alpha & - e^{-\phi_0} \sin \alpha
+ \chi_0 \cos \alpha\\
\sin \alpha & \cos \alpha\end{array}\right)\,e^{\phi_0/2},
\label{eq:pdlamb}
\end{equation}
Here $\alpha$ is an arbitrary parameter which will be fixed from the 
given transformed charges.

A classical NSNS superfivebrane, carrying an arbitrary classical charge  
$P_{(p_1, p_2)} = \Delta_{(p_1, p_2)}^{1/2} P_0$, 
with $\Delta_{(p_1, p_2)}$ an as yet undetermined dimensionless
factor and $P_0$ the charge unit which may be taken as the quantized 
unit charge (for example, the fundamental NSNS
fivebrane tension $T_5$), is associated with the 3-form field 
strength $H^{(1)}$. 
The meaning of the subscript $(p_1, p_2)$ will become clear when we 
discuss the charge quantization for the fivebranes. 
 The general classical fivebrane configuration which we will construct 
requires both the 3-form
field strengths $H^{(1)}$ and $H^{(2)}$ to be non-zero. Associated with 
this configuration is an arbitrary but given classical
charge doublet 
\begin{equation}
P = \left(\begin{array}{c} P^{(1)}\\ P^{(2)}\end{array}\right).
\label{eq:gcharged}
\end{equation}
{}From $P = (\Lambda^T)^{-1} \left(\begin{array}{c} P_{(p_1, p_2)}\\ 0
\end{array}\right)$ with $\Lambda$ given by
 Eq.~(\ref{eq:pdlamb}) and 
the relation $\cos^2 \alpha + \sin^2 \alpha = 1$, 
we  completely determine 
the SL(2,R) matrix  
\begin{equation}
\Lambda =\frac{1}{\Delta_{(p_1,p_2)}^{1/2}}
\left(\begin{array}{cc} 
e^{-\phi_0} P^{(1)}/P_0 &   - (P^{(2)}/P_0 + \chi_0 P^{(1)}/P_0) \\
+ \chi_0 e^{\phi_0} (P^{(2)}/P_0 + \chi_0 P^{(1)}/P_0)& + \chi_0 P^{(1)}/P_0  \\
&\\
e^{\phi_0} (P^{(2)}/P_0 + \chi_0 P^{(1)}/P_0)  & P^{(1)}/P_0\end{array}\right),
\label{eq:cdlamb}
\end{equation}
and the  $\Delta_{(p_1, p_2)}$ factor as, 
\begin{equation}
\begin{array}{rcl}
\Delta_{(p_1, p_2)}  &=&  e^{-\phi_0} \left(P^{(1)}/P_0 \right)^2 + \left(P^{(2)}/P_0 +
 \chi_0 P^{(1)}/P_0 \right)^2
 e^{\phi_0} \\
   &=& \left(P^{(1)}/P_0, P^{(2)} /P_0 \right) {\cal M}_0 \left(\begin{array}{c}
P^{(1)}/P_0 \\ P^{(2)}/P_0\end{array}\right),
\end{array} 
\label{eq:ddelta}
\end{equation}
where  ${\cal M}_0 = \left(\begin{array}{cc} \chi_0^2 + 
e^{-2\phi_0} & \chi_0\\ \chi_0 & 1\end{array}\right)\,e^{\phi_0}$.  
It is clear that as $\Delta_{(p_1, p_2)}$ in (11) is SL(2,R) invariant, the
charge $P_{(p_1, p_2)} = \Delta_{(p_1, p_2)}^{1/2} P_0$ is also 
SL(2,R) invariant. 

By now we have constructed the most general SL(2,R) fivebrane configuration carrying classical charges given by the charge 
doublet. The central charge (therefore the ADM mass per unit fivebrane which
can be calculated following \cite{lu} and the  tension measured in Einstein
 metric)
associated with this fivebrane is 
$P_{(p_1, p_2)} = \Delta_{(p_1, p_2)}^{1/2} P_0$ with 
$\Delta_{(p_1,p_2)}$ given by Eq.~(\ref{eq:ddelta}). The metric continues to be given by the metric in 
Eq.~(\ref{eq:nsfiveb}) but
now with $P = P_{(p_1, p_2)}$. The 3-form field strength doublet is simply
\begin{equation}
\begin{array}{rcl}
{\cal H} &=& (\Lambda^T)^{-1} \left(\begin{array}{c} P_{(p_1, p_2)}\\ 0
\end{array}\right)\epsilon_3\\
&=& \left(\begin{array}{c} P^{(1)}\\P^{(2)}\end{array}\right) \epsilon_3,
 \end{array}
\label{eq:t3form}
\end{equation}
as expected. The complex scalar is now
\begin{equation}
\begin{array}{rcl}
\lambda &=& \frac{a \left(i e^{-\phi}\right) + b}{c \left(i e^{-\phi}\right)
 + d}\\
&=& \frac{\chi_0 \Delta_{(p_1, p_2)} A_{(p_1, p_2)} + P^{(1)} P^{(2)}/P_0^2
\left(A_{(p_1, p_2)} - 1\right)  e^{- \phi_0} + i \Delta_{(p_1, p_2)} 
A_{(p_1, p_2)}^{1/2} 
e^{-\phi_0}}{\left(P^{(1)}/P_0\right)^2 
e^{-\phi_0} + A_{(p_1, p_2)} e^{\phi_0} \left(
\chi_0 P^{(1)}/P_0 + P^{(2)}/P_0\right)^2},
\end{array}
\label{eq:tlambda}
\end{equation}
 where $A_{(p_1, p_2)} = \left(1 + \frac{P_{(p_1, p_2)}}{2\rho^2}\right)^
{-1}$. Note that asymptotically as $\rho \rightarrow \infty$, $A_{(p_1,
p_2)} \rightarrow 1$ and therefore, $\lambda \rightarrow \lambda_0$ 
as expected. The real and imaginary part of ~(\ref{eq:tlambda}) give the transformed value
of the RR scalar and the dilaton of the theory. Partial results of the classical fivebrane was also obtained by Bergshoeff et
al.\cite{berbo}.

Our general classical fivebrane 
solution also preserves half of the spacetime supersymmetry as the original 
pure NSNS fivebrane since the global SL(2,R) 
transformation commutes with the supersymmetry transformation. 
Therefore, our general fivebrane solution continues to be BPS which implies 
that the ADM mass per  unit fivebrane volume,
the central charge $P_{(p_1, p_2)}$ and the fivebrane tension measured in Einstein metric are all the same in proper units.

So far we have only considered the most general classical type IIB fivebrane solution preserving half of the spacetime 
supersymmetry in the sense that the two charges $P^{(1)}$ and $P^{(2)}$ can be arbitrary. As mentioned before, due to the
existence of both F-string and D-string, the magnetic duals of the NSNS fivebrane and RR fivebrane, respectively, each of
the two charges must be quantized separately in terms of the 
unit charge $P_0$. For example, F-string whose charge
doublet corresponds to $Q^{(1)} = Q_0, Q^{(2)} = 0$ implies that the 
charge $P^{(1)}$ is quantized in terms of the unit
charge $P_0 = 2\pi / Q_0$. So the charge doublet for a general quantum 
mechanically allowable fivebrane solution is
\begin{equation}
P = \left(\begin{array}{c} p_1\\p_2\end{array}\right) P_0,
\label{eq:qcharge}
\end{equation}
where $p_1, p_2$ are two integers. In terms of the unit charge $P_0$, the charge doublet should remain to be an integral
doublet under quantum-mechanically allowable transformation. This necessarily breaks the continuous SL(2,R) symmetry of type
IIB supergravity to a discrete SL(2,Z) of type IIB non-perturbative string theory whose elements take only integral values.

The most general quantum-mechanically allowable fivebrane 
configuration can be obtained simply by setting 
$P^{(1)} = p_1 P_0, P^{(2)} = p_2 P_0$ in the above classical 
fivebrane solution. For example, 
\begin{equation}
\Delta_{(p_1, p_2)} = \left(p_1, p_2\right) {\cal M}_0 \left(\begin{array}{c}
p_1 \\ p_2\end{array}\right),
\label{eq:qdelta}
\end{equation}
which is now SL(2,Z) invariant. Therefore, the ADM mass per unit fivebrane volume $M_{(p_1, p_2)}$, the central charge
$P_{(p_1, p_2)}$ and the fivebrane tension $T_{(p_1, p_2)}$ measured in Einstein metric are all SL(2,Z) invariant. In proper
units, we can set all three equal in which case we can take $P_0$ 
as the fundamental NSNS fivebrane tension 
$T_5$. Then
for a $(p_1, p_2)$-fivebrane, we have  
\begin{equation}
\begin{array}{rcl}
M_{(p_1, p_2)} &=& P_{(p_1, p_2)} = \Delta_{(p_1, p_2)}^{1/2} T_5\\
               &=& \sqrt{e^{-\phi_0} p_1^2 + \left(p_2 + p_1 \chi_0\right)^2 
e^{\phi_0}} 
~T_5,
\end{array}
\label{eq:mct}
\end{equation}
The $(p_1, p_2)$-fivebrane tension measured in type IIB string metric  is
\begin{equation}
T_{(p_1, p_2)} = g^{- 3/2}_s \sqrt{g_s^{-1} p_1^2 + \left(p_2 +
 p_1 \chi_0\right)^2 g_s }~ T_5,
\label{eq:stension}
\end{equation}
where $g_s = e^{\phi_0}$ is the string coupling. Here $T_{(p_1, p_2)}$ 
gets scaled by $g_s^{-3/2}$
because it has dimensionality of $({\rm length})^{-6}$. 
{}From the above tension formula, we can see that 
the tension for a (1, 0)-fivebrane, i.e., 
a NSNS fivebrane, is equal to $T_5/g_s^2$ and the tension for a 
(0, 1)-fivebrane, i.e., a RR fivebrane, is equal to $T_5/g_s$, 
as expected in both cases. Implementation of charge quantization on the
charges carried by the fivebrane consists of the second step of 
our construction.

The above $(p_1, p_2)$-fivebrane configuration encodes all the information 
about the SL(2,Z) multiplets of the type IIB
superfivebranes. Note that for given asymptotic values of the scalars, i.e., 
for a given vacuum, each of the infinitely
many integral doublets $(p_1, p_2)^T$ gives a different value for the 
$\Delta_{(p_1, p_2)}$ which cannot be related to each
other by a SL(2,Z) transformation since it is invariant by such a 
transformation. Further, this $\Delta_{(p_1, p_2)}$
measures the mass per unit fivebrane volume, the central charge and the 
tension. Therefore, we can use this factor to label
different SL(2,Z) multiplets. Within each such multiplet, we have a collection
 of infinitely many discrete vacua and a
collection of infinitely many integral charge doublets. Each of such vacua 
and its corresponding integral charge
doublet are obtained from the given initial vacuum $\lambda_0 = \chi_0 + 
i e^{-\phi_0}$ and the given initial charge doublet
$(p_1, p_2)^T$ by a particular SL(2,Z) transformation. Picking a special vaccum 
in such a multiplet will break the SL(2,Z)
symmetry spontaneously. In other words, all the fivebrane configurations in 
such a multiplet are physically equivalent. The
physically inequivalent fivebrane configurations are those with different
 $\Delta_{(p_1, p_2)}$ values which correspond to
different integral doublets $(p_1, p_2)^T$ for a fixed $\lambda_0$, i.e., a 
fixed vacuum. 

The third step of our construction consists of the determination of the 
condition for 
 which a $(p_1, p_2)$-fivebrane is absolutely
stable (For discussion of the stability of SL(2,Z) strings, see \cite{schtwo}).
 We have noted that the tension of a $(p_1, p_2)$-fivebrane is given by 
$T_{(p_1, p_2)} = \Delta_{(p_1, p_2)}^{1/2} 
T_5$ and it can easily be checked that the tensions satisfy the following 
triangle relation (also called ``tension gap"
 equation), irrespective of the vaccum
moduli,
\begin{equation}   
 T_{(p_1, p_2)} + T_{(q_1, q_2)} \geq T_{(p_1+q_1, p_2+q_2)},
 \label{eq:tensionie}
\end{equation}
 where the equality holds if and only if $p_1 q_2 = p_2 q_1$, i.e., when 
$p_1 = n q_1, p_2 = n q_2$ with $n$ being an
 integer (assuming $p_1 \geq q_1, p_2 \geq q_2$ without losing generality). 
So when $p_1$ and $p_2$ are relatively 
 prime, the tension inequality prevents the fivebrane state from decaying to 
fivebrane states with lower tensions or
 lower masses. Therefore such a $(p_1, p_2)$-fivebrane is absolutely stable. 
The same conclusion can also be drawn when
 the central charge triangle inequality relation (now called ``charge gap"
 equation) and the charge conservation
 are employed. We therefore finally conclude that all physically inequivalent 
absolutely stable BPS 
 $(p_1, p_2)$-fivebrane configurations are those corresponding to all possible 
integral doublets $(p_1, p_2)^T$ with
 $p_1, p_2$ coprime and with $\lambda_0$ belonging to the fundamental region
  of SL(2,Z), i.e., SL(2,Z)$\setminus$ SL(2,R)/SO(2).
 
 \section{Discussion} 
 
 We have demonstrated the procedure how to construct a 
general solution of a theory in detail when the theory has a global
 symmetry starting from a known particular solution of this theory, 
with the example of type IIB $(p_1, p_2)$-fivebranes. 
 This procedure can also be applied to construct U-duality p-branes 
in diverse 
dimensions from the known NSNS
 p-branes (for example, see \cite{dufkl}) whether they are 
supersymmetric or not
 as we did recently \cite{lurone}. We are carefully using the word 
``construct" rather than ``generate" as
 we cannot really generate all the solutions from a given one simply 
by applying the underlying global symmetry
 transformations (since these transformations will in general 
preserve the corresponding masses, charges and tensions) just as 
 we cannot relate different particles with different masses and spins 
simply by Lorentz transformations. In order to
 generate solutions from a given one, additional 
``symmetries" are needed which are most likely symmetries of certain spectra
 (for example, BPS states)
 of the underlying theory rather than symmetries of the theory itself 
as discussed recently by 
 L\"{u} et al. \cite{lps}. 
 
 Finally it should be pointed out that we could determine the 
SL(2,R) matrix $\Lambda$~(\ref{eq:cdlamb}) completely in terms of 
asymptotic values 
 $\chi_0$ and $\phi_0$, and the charges $P^{(1)}$ and $P^{(2)}$ 
because we have four parameters (three for
 SL(2,R) matrix elements and one for the $\Delta_{(p_1, p_2)}$ factor) 
and four equations (two real equations for the
 asymptotic complex scalars and two for the charges). However, 
this is not true in general. For example, for other U-duality
 groups, the number  of group matrix parameters plus the corresponding 
$\Delta$ factor exceeds that of the available equations. 
 At first sight, it appears that the general solution thus constructed 
contains certain arbitrary parameters. As
 demonstrated recently by the present authors \cite{lurone}, this never happens. The $\Delta$ factor is always completely determined in terms of
 the asymptotic values of scalars and the charges. The undetermined 
group elements never enter the quantities which characterize
 the solution such as
 scalars and field strengths. In other words, the general solution 
is completely determined by the charges and asymptotic
 values of scalars as expected.

\section*{Acknowledgments}
JXL is grateful to the organizers to allow him to speak in the ICTP conference of superfivebranes and
physics in 5 + 1 dimensions and acknowledges the support of NSF Grant PHY-9722090.

\end{document}